\documentclass[10pt,conference]{IEEEtran}
\IEEEoverridecommandlockouts

\newcommand*\rot{\rotatebox{90}}

\usepackage[table,xcdraw]{xcolor}
\usepackage{cite}
\usepackage{amsmath,amssymb,amsfonts}
\usepackage{algorithmic}
\usepackage{graphicx}
\usepackage{svg}
\usepackage{lipsum} 
\usepackage{textcomp}
\usepackage[skins]{tcolorbox}
\usepackage{dblfloatfix}
\usepackage{balance}

\def\BibTeX{{\rm B\kern-.05em{\sc i\kern-.025em b}\kern-.08em
    T\kern-.1667em\lower.7ex\hbox{E}\kern-.125emX}}
\begin{document}

\title{Blockchain Developer Experience: A Multivocal Literature Review
}

\author{
\IEEEauthorblockN{Pamella Soares\textsuperscript{1}, Allysson Allex Araújo\textsuperscript{2}, Giuseppe Destefanis\textsuperscript{3}, \\Rumyana Neykova\textsuperscript{3}, Raphael Saraiva\textsuperscript{1}, Jerffeson Souza\textsuperscript{1}}
\IEEEauthorblockA{\textsuperscript{1}\textit{Graduate Program in Computer Science}, \textit{State University of Ceará}, Fortaleza, Brazil \\
\textsuperscript{2}\textit{Science and Technology Center}, \textit{Federal University of Cariri}, Juazeiro do Norte, Brazil \\
\textsuperscript{3}\textit{Department of Computer Science}, \textit{Brunel University}, London, United Kingdom}
}

\maketitle

\begin{abstract}
The rise of smart contracts has expanded blockchain’s capabilities, enabling the development of innovative decentralized applications (dApps). However, this advancement brings its own challenges, including the management of distributed architectures and immutable data. Addressing these complexities requires a specialized approach to software engineering, with blockchain-oriented practices emerging to support development in this domain. Developer Experience (DEx) is central to this effort, focusing on the usability, productivity, and overall satisfaction of tools and frameworks from the engineers’ perspective. Despite its importance, research on Blockchain Developer Experience (BcDEx) remains limited, with no systematic mapping of academic and industry efforts. To bridge this gap, we conducted a Multivocal Literature Review analyzing 62 to understand the distribution of BcDEx sources, practical implementations, and their impact. Our findings revealed that academic focus on BcDEx is limited compared to the coverage in gray literature, which primarily includes blogs (41.8\%) and corporate sources (21.8\%). Particularly, development efficiency, multi-network support, and usability are the most addressed aspects in tools and frameworks. In addition, we found that BcDEx is being shaped through five key perspectives: complexity abstraction, adoption facilitation, productivity enhancement, developer education, and BcDEx evaluation.
\end{abstract}


\section{Introduction}
\label{sec:intro}

Blockchain technology emerged in 2008 with Bitcoin \cite{nakamoto2009bitcoin} as a pioneer in electronic peer-to-peer currency transactions without intermediary authorities that use a consensus protocol based on cryptographic challenges \cite{bashir2017mastering}. This technology ensures consensus in decentralized networks and guarantees transaction auditability, authenticity, availability, and integrity~\cite{greve2018blockchain}. In particular, adopting smart contracts by Ethereum \cite{buterin2013ethereum} expanded the capabilities beyond financial transactions and enabled programmable business logic executed as transactions on the blockchain \cite{xu2019blockchain}. This component has enabled the diffusion of Decentralized Applications (dApps), which operate independently of a central entity, utilizing blockchain for both storage and processing \cite{metcalfe2020ethereum}. 

Blockchain development presents different challenges due to its distributed nature and high operational costs~\cite{li2021services, kotha2023complexity}. Unlike traditional software, dApp engineering requires practices tailored to immutable databases, peer-to-peer networks, and novel security mechanisms~\cite{kassab2021blockchain}. In light of these particularities, Blockchain-Oriented Software Engineering seeks to shape directions for effective software development for blockchain, serving as a bridge between conventional software engineering and these particular technical constraints \cite{porru2017blockchain, destefanis2018smart}.

Multiple development environments and a steep learning curve characterize the blockchain development ecosystem. Hence, there is a continual need for new approaches to enhance Developer Experience (DEx), as improvements in developer efficiency, security, and productivity can directly influence the success of system implementations \cite{razzaq2024systematic, nylund2020multivocal}. According to Fagerholm and M{\"u}nch~\cite{fagerholm2012developer}, DEx encompasses the experiences related to all types of artifacts and activities that a developer encounters in the software development. Thus, we can observe a strong connection between DEx and the cooperative and human aspects of software engineering.

Despite growing attention to DEx among SE scholars and practitioners \cite{razzaq2024systematic, greiler2022actionable}, research specifically addressing Blockchain Developer Experience (BcDEx) remains considerably limited. In fact, there are no systematic mappings that capture how academic and professional communities approach aspects of BcDEx. This gap is relevant, as the inherent complexity and rapid evolution of blockchain technologies emphasize the need to understand and enhance developers’ interactions with these tools. Addressing this issue could contribute to improving developer productivity and satisfaction, directly impacting the quality, security, and adoption of dApps.

Based on the motivation previously discussed, this study conducts a Multivocal Literature Review (MLR)~\cite{garousi2019guidelines} to analyze the distribution of BcDEx literature sources, practical implementations, and their impact on BcDEx. In particular, MLR is considerably suited to this research, as it combines insights from both academic and gray literature (such as blogs, corporate publications, and technical documentation), providing a broader and up-to-date perspective on BcDEx. To reach our research objective, we define the following research questions (RQs).

\begin{itemize} 
\item \textbf{RQ1) What is the distribution and nature of academic and industry sources related to BcDEx?} \textit{Rationale:} Analyze the distribution of publications between White Literature (WL) and Gray Literature (GL) related to BcDEx, identifying types of publications, venues, research approaches, and trends over the years.

\item \textbf{RQ2) What categories of practical sources related to BcDEx have been discussed in the literature?} \textit{Rationale:} Identify and categorize practical sources that influence BcDEx, analyzing key features and capabilities. 

\item \textbf{RQ3) In what ways have the sources discussed in the literature been shaping the BcDEx in practice?} \textit{Rationale:} 
Examine how the sources shape the BcDex in practice, focusing on their effects and benefits.
\end{itemize} 

Therefore, this research offers contributions to both SE academia and industry. For academia, it bridges the gap between industry practices and academic research by systematizing BcDEx insights from industry sources on blockchain development tools, practices, and challenges. We also highlight research opportunities, particularly in empirical validation studies and the creation of robust methods for BcDEx assessment. For industry practitioners, we provide a structured categorization of current context of BcDEx, focusing on critical aspects to guide blockchain development teams in making informed decisions.

\section{Methodological Procedures} 
\label{sec:method}

\begin{figure}[!h]
\centerline{\includegraphics[scale=0.51]{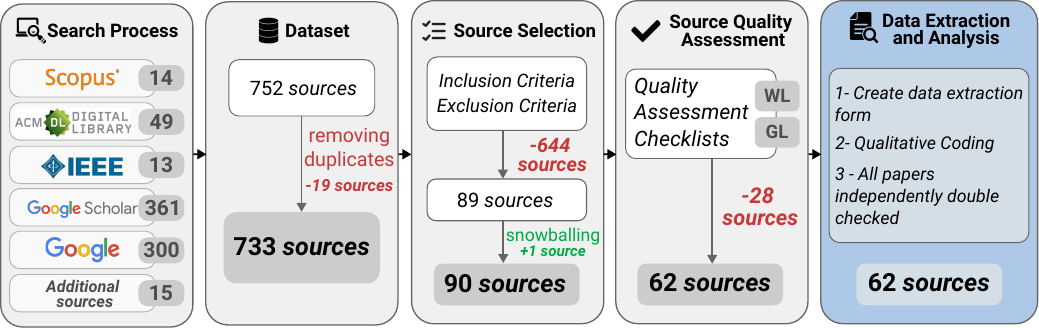}}
\caption{Multivocal Literature Review process.}
\label{fig:method}
\end{figure}

We conducted our Multivocal Literature Review (MLR) following the guidelines provided by Garousi et al.~\cite{garousi2019guidelines}. This method considers theoretical and practical development aspects by including WL and GL. We use this approach as it covers contemporary topics and allows us to understand industry aspects, where we can often find up-to-date resources. Figure~\ref{fig:method} illustrates how our process was organized into five stages. In the following, we describe each one of these stages.

\subsubsection{\textbf{Search Process}}
Initially, we conducted this step using Scopus, IEEE Xplore, ACM Digital Library as digital libraries, and Google Scholar as databases to find studies for WL. These libraries were selected based on their coverage, search capabilities, and access to full texts. To obtain sources from GL, we conducted searches using Google, a web search engine commonly used in GL reviews. We conducted pilot searches using basic ``blockchain" and ``developer experience" terms. We identified additional relevant keywords from these preliminary searches and expanded our search terms. The final search was performed in June 2024 using the following terms:

\begin{table}[ht!]
\centering
\resizebox{0.48\textwidth}{!}{%
\begin{tabular}{|l|}
\hline
\footnotesize
\begin{tabular}[c]{@{}l@{}}(``blockchain" OR ``smart contract") AND (``developer experience" \\OR ``developer usability" OR ``software engineer experience" OR \\``programmer experience" OR ``DevX" OR ``DEx")\end{tabular} \\ \hline
\end{tabular}}
\end{table}

Regarding the WL, we carried out this search using the title, abstract, and keywords of the studies in most databases, except Google Scholar. We applied the search string to the full text when the database did not support searches using these fields. For WL, the initial search returned 437 studies. As Garousi et al.~\cite{garousi2019guidelines} suggested, we used an `effort-bounded' mechanism as the stopping criterion for GL, including only the top $N$ search engine results. In this regard, we selected the top 300 search results. During the initial search process, we find several relevant platforms for blockchain development. However, some of the most widely used and recognized tools in the developer community were not identified due to the inherent limitations of automated searches. 

Based on the authors' practical experience in developing dApps using established tools (e.g., Infura, and Alchemy), we searched for alternative solutions in the market, prioritizing those with greater adoption by the developer community and relevance to BcDEx, resulting in the inclusion of 15 additional technical sources. These sources were subjected to the same evaluation processes and quality criteria as the others. Considering both literatures, the process yielded a total of 752 sources. After removing duplicates, we obtained 733 sources.

\subsubsection{\textbf{Source Selection}}
In the this step, we established Inclusion Criteria (IC) and Exclusion Criteria (EC) to filter and select relevant sources. The ICs are as follows: (IC1) sources must be written in English; (IC2) sources must focus on blockchain and mention or discuss DEx in this context; (IC3) sources must evaluate DEx factors (or synonyms) in blockchain solutions; and (IC4) sources must present solutions that may impact DEx factors (or synonyms). Given the broad scope of DEx, we considered sources that relate any DEx dimension proposed by Fagerholm and M{\"u}nch~\cite{fagerholm2012developer}: development infrastructure (Cognition), feelings about the work (Affect), and the value of the contribution itself (Conation). We included sources referencing ``developer experience'' explicitly and implicitly. In the latter case, we include developer-focused solutions with sentences such as \textit{``query blockchain data with two lines of code''} or \textit{``simplify account abstraction development''}, which can impact aspects of DEx.

The ECs were defined as follows: (EC1) sources without full-text availability; (EC2) secondary or tertiary papers (e.g., systematic literature reviews, surveys, etc.); (EC3) papers in the form of editorials, proceedings, etc., as they do not provide sufficient information; and (EC4) sources where the topic of ``developer experience" is only mentioned without any relation to BcDEx or the solution presented. We applied the selection criteria through a peer review process. Specifically, the first author applied the ECs and ICs, while another coauthor conducted a double-check afterward. In cases of conflict, the researchers met virtually with other coauthors to reach a consensus. After applying ICs and ECs, we selected 89 relevant sources. In addition, we employed the ``snowballing'' technique on the set of papers from the WL. This method involves following citations backward or forward from a set of seed papers~\cite{garousi2019guidelines}. As a result of this process, only one paper was selected from snowballing forward.

\subsubsection{\textbf{Source Quality Assessment}}
We also assessed sources to determine their quality in WL and GL, applying a checklist with four questions used by Martin and Runeson~\cite{host2007checklists} and Baysal et al.~\cite{baysal2023blockchain}. This process was similar to the selection process, in which two authors evaluated each source based on checklist questions. For this purpose, we used a checklist with a 3-point scale (0 -- No, 1 -- Partly, 2 -- Yes). We calculated the average of both reviewer scores and finally included review sources with a rating of 0.7 or higher from 0 to 1. Table \ref{tab:quality} shows the quality assessment checklists of WL and GL, respectively. As a result, we rejected 28 sources, resulting in 62 for the final set of primary studies.

\begin{table}[h]
\caption{Quality assessment Checklists}
\label{tab:quality}
\renewcommand{\arraystretch}{1.08}
\resizebox{0.48\textwidth}{!}{%
\begin{tabular}{ll}
\hline
\multicolumn{2}{l}{\textbf{WL Quality Assessment Checklist}} \\ \hline
\textbf{Q1} & Are the authors' intentions with the research made clear? \\
\textbf{Q2} & \begin{tabular}[c]{@{}l@{}}Does the study contain conclusion, implications for practice \\ and future research?\end{tabular} \\
\textbf{Q3} & Does the study give a realistic and credible impression? \\
\textbf{Q4} & Are the challenges or solutions adequately defined in detail? \\ \hline
\multicolumn{2}{l}{\textbf{GL Quality Assessment Checklist}} \\ \hline
\textbf{Q1} & Does the source have a clearly aim? \\
\textbf{Q2} & Does the source have a clearly stated date? \\
\textbf{Q3} & Does the source give a realistic and credible impression? \\
\textbf{Q4} & Are the challenges or solutions adequately defined in detail? \\ \hline
\end{tabular}}
\end{table}
 
\subsubsection{\textbf{Data Extraction and Analysis}}
We aimed to extract and analyze the available data based on the RQs defined earlier. The data extraction process was systematically designed to address each RQ specifically. The data were then analyzed quantitatively using descriptive statistics and qualitatively through  coding~\cite{miles1994qualitative}. To ensure reliability in the qualitative analysis, the first author independently performed the coding process, and subsequently, a second co-author performed validation reviewing each code and its associated evidence from the sources. When disagreements emerged, both authors discussed until consensus or consulted a third co-author for mediation. Additionally, two other co-authors reviewed the final list of codes.

To answer \textbf{RQ1}, we conducted quantitative and qualitative analyses. For GL, we extracted the source type (blog, video, wiki, documentation, etc.) and publication platform. For WL, we identified authors' affiliation (academic, industry, or collaboration), publication type (conference, journal, etc.), and venue. For the WL's qualitative analysis, we applied a deductive open coding approach with predefined categories based on Wieringa et al.~\cite{wieringa2006requirements}, which defines six types of research: solution proposal, validation research, evaluation research, personal experience paper, philosophical paper, and opinion paper. We conducted a descriptive quantitative analysis of the distribution of primary sources, based on the collected data regarding their type and nature.

To answer \textbf{RQ2}, we employed a combined deductive and inductive coding approach. Our analysis consisted of two main steps. First, using deductive coding, we classified each source into one of the following groups:

\begin{itemize}
\item \textbf{[1st Group] Tool, platform/service, language, method/technique, model, process, or framework:} We extracted the \textit{main features} presented by the source or the \textit{functioning} of proposed approaches.

\item \textbf{[2nd Group] Heuristic/guidelines, empirical results only, or other:} We extracted the \textit{main topics} approached or discussed by the source.
\end{itemize}

We divided them in these two groups to gain an understanding of both the tangible functionalities and the general thematic discussions that influence the BcDex. Then, following Garousi et al.\cite{garousi2019guidelines}, we applied inductive coding for each group where factors emerged from iterative ``open" and ``axial" coding. This approach allowed us to extract specific data from each category. For instance, in the first group, we classified features according to their capabilities - such as \textit{``deploying a contract with a single line of command"}, which was categorized under ``Efficiency of Development". Based on our analysis, we identified 10 main capability categories: (i) Efficiency of Development; (ii) User Interface (UI) and Usability; (iii) Reporting and Analytics; (iv) Quality Assurance; (v) Multi-Network Support; (vi) Performance and Scalability; (vii) Privacy and Security; (viii) Specification and Documentation; (ix) Asset Management; and (x) Storage. 

We classified the sources from the second group from a broader perspective considering how the content and topics were addressed. For Heuristics/Guidelines, we noticed some blogs compared Web3 tools or Blockchain platforms, for example. Thus, we classified it as a ``Tool Comparison". In this context, we identified that the sources were usually related to presenting best practices, new protocols and blockchains, emerging trends and new features, tool comparisons, empirical results or surveys, and personal experiences.

Regarding \textbf{RQ3}, we employed an inductive open coding approach to extract aspects that demonstrate how the sources influence BcDEx, including (i) approaches to abstract complexity and enhance usability, (ii) strategies for facilitating adoption, (iii) impacts on developer productivity and workflow, (iv) educational and support initiatives, and (v) empirical evaluations of BcDEx. The complete mapping between sources and its categorization is available in our repository\cite{repository_mlr_bcdex}.

\section{Results}
\label{sec:results}

\subsection{RQ1) What is the distribution and nature of academic and industry sources related to BcDEx?} 
\label{sec:rq1}

\begin{figure}[htbp]
\centerline{\includegraphics[scale=0.8]{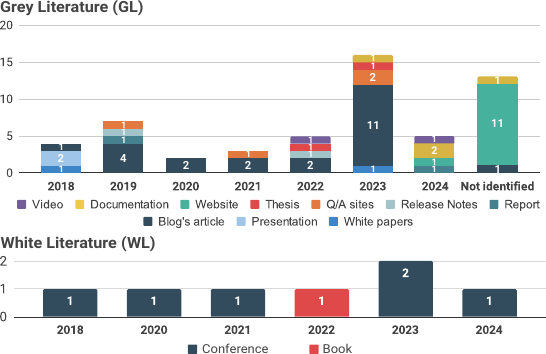}}
\caption{Distribution of WL and GL type sources over the years.}
\label{fig:publication-type}
\end{figure}

Figure \ref{fig:publication-type} presents the distribution of selected primary sources considering publication type for WL and source type for GL and their respective publication years. As we can see, the number of WL studies (7) is considerably smaller than the GL sources (55). This finding reveals a gap in academic research explicitly focusing on BcDEx studies. Only one study was published yearly, except in 2023, with two publications. On the other hand, GL shows progress in discussions about BcDEx, which is driven by industry efforts to enhance resources for advancing blockchain solutions. Similar to WL, GL's results also increased in publications in 2023.

As for the publication type and venues in WL, seven studies were published in the following conferences: IEEE International Conference on Data Mining Workshops [WL01], International Conference on Evaluation and Assessment in Software Engineering [WL02], Formal Methods in Computer-Aided Design Conference [WL03], Workshop in Blockchain: Theory, Technologies and Applications [WL07], International Conference on Blockchain Computing and Applications [WL04], and International Conference on Software Engineering [WL06]. Springer published the book in the ``Optimization and Its Applications'' series [WL05]. Regarding the authors' affiliations in WL, we found that authors were predominantly from academia (4 studies), with only one industry-authored study and one industry-academia collaboration.

In contrast, the GL literature is predominantly represented by blog articles (41.8\%), followed by websites (21.8\%), Q/A sites, and documentation (7.3\%) each. Other sources (videos, white papers, presentations, reports, release notes, and thesis) each comprise 3.6\%. We could not identify the publication date for 12 sources, primarily websites related to blockchain products. Approximately 52.63\% of the sources are published by the enterprises themselves, offering relevant guidance for developers. Finally, regarding the results of research types in WL literature, we found that three studies [WL01, WL05, WL06] were classified as `Solution Proposal', which typically involve solution techniques without a ``full-blown validation''. In addition, the other four studies [WL02, WL03, WL04, WL07] were categorized as `Validation Research' as they utilize a methodologically sound set of methods such as experiments, simulation, prototyping, mathematical analysis, and mathematical proof of properties. However, we observed limited empirical evaluations in WL and GL, with minimal systematic research exploring DEx and the practical impact of blockchain development tools and techniques in real-world contexts.

\begin{tcolorbox}[right=0.1cm,left=0.1cm,top=0.1cm,bottom=0.1cm]
\small\textbf{Answer to RQ1:} \textit{Our findings show that the BcDEx knowledge ecosystem is predominantly driven by industry sources within GL, with blog articles (41.8\%) and company websites (21.8\%) offering practical guidance to developers. In contrast, we identified only seven studies in the WL, mostly classified as Solution Proposals and Validation Research, highlighting a research gap concerning empirical evaluations.}
\end{tcolorbox}

\subsection{RQ2) What categories of practical sources related to BcDEx have been discussed in the literature?}
\label{sec:rq2}
For this analysis, we identified and categorized practical sources influencing BcDEx, examining their key features and capabilities. Table \ref{tab:final-set} presents the findings of the sources and their respective categories, which are discussed in detail below.

\begin{table}[!h]
{\fontsize{5.4}{5.4}\selectfont
\renewcommand{\arraystretch}{1.08}
\setlength{\tabcolsep}{3.3pt}
\arrayrulecolor[rgb]{0.8,0.8,0.8}
\caption{MLR Final Source Set.}
\label{tab:final-set}
\begin{tabular}{|c|c|c|l|llllllllll|}
\hline
 &  &  & \multicolumn{1}{c|}{\textbf{\begin{tabular}[c]{@{}c@{}}{[}2nd Group{]}\\ Discussed topics\end{tabular}}} & \multicolumn{10}{c|}{\textbf{\begin{tabular}[c]{@{}c@{}}{[}1st Group{]}\\ Main features\end{tabular}}} \\ \cline{4-14} 
{\textbf{\rot{Literature Type}}} & {\textbf{\rot{Reference}}} & {\textbf{\rot{Year}}} & \multicolumn{1}{c|}{\textbf{--}} & \multicolumn{1}{c|}{\textbf{\rot{Efficiency of Development}}} & \multicolumn{1}{c|}{\textbf{\rot{UI and Usability}}} & \multicolumn{1}{c|}{\textbf{\rot{Reporting and Analytics}}} & \multicolumn{1}{c|}{\textbf{\rot{Quality Assurance}}} & \multicolumn{1}{c|}{\textbf{\rot{Multi-Network Support}}} & \multicolumn{1}{c|}{\textbf{\rot{Performance and Scalability}}} & \multicolumn{1}{c|}{\textbf{\rot{Privacy and Security}}} & \multicolumn{1}{c|}{\textbf{\rot{Spec. and Documentation}}} & \multicolumn{1}{c|}{\textbf{\rot{Asset Management}}} & \multicolumn{1}{c|}{\textbf{\rot{Storage}}} \\ \hline
WL01 & \cite{wl01_lu2018design} & 2018 &  & \multicolumn{1}{l|}{\cellcolor[HTML]{D9EAD3}{\color[HTML]{D9EAD3} 1}} & \multicolumn{1}{l|}{} & \multicolumn{1}{l|}{} & \multicolumn{1}{l|}{} & \multicolumn{1}{l|}{} & \multicolumn{1}{l|}{\cellcolor[HTML]{D9EAD3}{\color[HTML]{D9EAD3} 1}} & \multicolumn{1}{l|}{\cellcolor[HTML]{D9EAD3}{\color[HTML]{D9EAD3} 1}} & \multicolumn{1}{l|}{} & \multicolumn{1}{l|}{} &  \\ \hline
WL02 & \cite{wl02_tschannen2020evaluation} & 2020 &  & \multicolumn{1}{l|}{} & \multicolumn{1}{l|}{} & \multicolumn{1}{l|}{} & \multicolumn{1}{l|}{\cellcolor[HTML]{D9EAD3}{\color[HTML]{D9EAD3} 1}} & \multicolumn{1}{l|}{} & \multicolumn{1}{l|}{} & \multicolumn{1}{l|}{} & \multicolumn{1}{l|}{} & \multicolumn{1}{l|}{} &  \\ \hline
WL03 & \cite{wl03_dharanikota2021celestial} & 2021 &  & \multicolumn{1}{l|}{} & \multicolumn{1}{l|}{\cellcolor[HTML]{D9EAD3}{\color[HTML]{D9EAD3} 1}} & \multicolumn{1}{l|}{} & \multicolumn{1}{l|}{} & \multicolumn{1}{l|}{} & \multicolumn{1}{l|}{} & \multicolumn{1}{l|}{} & \multicolumn{1}{l|}{\cellcolor[HTML]{D9EAD3}{\color[HTML]{D9EAD3} 1}} & \multicolumn{1}{l|}{} &  \\ \hline
WL04 & \cite{wl04_morhavc2023paraspell} & 2023 &  & \multicolumn{1}{l|}{\cellcolor[HTML]{D9EAD3}{\color[HTML]{D9EAD3} 1}} & \multicolumn{1}{l|}{\cellcolor[HTML]{D9EAD3}{\color[HTML]{D9EAD3} 1}} & \multicolumn{1}{l|}{} & \multicolumn{1}{l|}{} & \multicolumn{1}{r|}{\cellcolor[HTML]{D9EAD3}{\color[HTML]{D9EAD3} 1}} & \multicolumn{1}{l|}{} & \multicolumn{1}{l|}{} & \multicolumn{1}{l|}{} & \multicolumn{1}{l|}{\cellcolor[HTML]{D9EAD3}{\color[HTML]{D9EAD3} 1}} &  \\ \hline
WL05 & \cite{wl05_mcconaghy2022ocean} & 2022 &  & \multicolumn{1}{l|}{\cellcolor[HTML]{D9EAD3}{\color[HTML]{D9EAD3} 1}} & \multicolumn{1}{l|}{} & \multicolumn{1}{l|}{} & \multicolumn{1}{l|}{} & \multicolumn{1}{l|}{} & \multicolumn{1}{l|}{} & \multicolumn{1}{l|}{\cellcolor[HTML]{D9EAD3}{\color[HTML]{D9EAD3} 1}} & \multicolumn{1}{l|}{} & \multicolumn{1}{l|}{\cellcolor[HTML]{D9EAD3}{\color[HTML]{D9EAD3} 1}} &  \\ \hline
WL06 & \cite{wl06_van2024verifying} & 2024 &  & \multicolumn{1}{l|}{\cellcolor[HTML]{D9EAD3}{\color[HTML]{D9EAD3} 1}} & \multicolumn{1}{l|}{} & \multicolumn{1}{l|}{} & \multicolumn{1}{l|}{\cellcolor[HTML]{D9EAD3}{\color[HTML]{D9EAD3} 1}} & \multicolumn{1}{l|}{} & \multicolumn{1}{l|}{} & \multicolumn{1}{l|}{} & \multicolumn{1}{l|}{} & \multicolumn{1}{l|}{} &  \\ \hline
WL07 & \cite{wl07_velasco2023evaluation} & 2024 &  & \multicolumn{1}{l|}{} & \multicolumn{1}{l|}{} & \multicolumn{1}{l|}{} & \multicolumn{1}{l|}{} & \multicolumn{1}{l|}{} & \multicolumn{1}{l|}{} & \multicolumn{1}{l|}{} & \multicolumn{1}{l|}{\cellcolor[HTML]{D9EAD3}{\color[HTML]{D9EAD3} 1}} & \multicolumn{1}{l|}{} &  \\ \hline
GL01 & \cite{gl01_vasconcelos} & 2022 & \multicolumn{1}{c|}{\begin{tabular}[c]{@{}c@{}}Empirical results or surveys\end{tabular}} & \multicolumn{1}{l|}{} & \multicolumn{1}{l|}{} & \multicolumn{1}{l|}{} & \multicolumn{1}{l|}{} & \multicolumn{1}{l|}{} & \multicolumn{1}{l|}{} & \multicolumn{1}{l|}{} & \multicolumn{1}{l|}{} & \multicolumn{1}{l|}{} &  \\ \hline
GL02 & \cite{gl02_pharr} & 2018 &  & \multicolumn{1}{l|}{\cellcolor[HTML]{D9EAD3}{\color[HTML]{D9EAD3} 1}} & \multicolumn{1}{l|}{} & \multicolumn{1}{l|}{} & \multicolumn{1}{l|}{} & \multicolumn{1}{l|}{} & \multicolumn{1}{l|}{} & \multicolumn{1}{l|}{} & \multicolumn{1}{l|}{\cellcolor[HTML]{D9EAD3}{\color[HTML]{D9EAD3} 1}} & \multicolumn{1}{l|}{} &  \\ \hline
GL03 & \cite{gl03_xandeum} & 2023 &  & \multicolumn{1}{l|}{} & \multicolumn{1}{l|}{} & \multicolumn{1}{l|}{} & \multicolumn{1}{l|}{} & \multicolumn{1}{l|}{} & \multicolumn{1}{l|}{} & \multicolumn{1}{l|}{} & \multicolumn{1}{l|}{} & \multicolumn{1}{l|}{} & \cellcolor[HTML]{D9EAD3}{\color[HTML]{D9EAD3} 1} \\ \hline
GL04 & \cite{gl04_barasti} & 2023 &  & \multicolumn{1}{l|}{\cellcolor[HTML]{D9EAD3}{\color[HTML]{D9EAD3} 1}} & \multicolumn{1}{l|}{} & \multicolumn{1}{l|}{} & \multicolumn{1}{l|}{} & \multicolumn{1}{l|}{\cellcolor[HTML]{D9EAD3}{\color[HTML]{D9EAD3} 1}} & \multicolumn{1}{l|}{} & \multicolumn{1}{l|}{} & \multicolumn{1}{l|}{\cellcolor[HTML]{D9EAD3}{\color[HTML]{D9EAD3} 1}} & \multicolumn{1}{l|}{} &  \\ \hline
GL05 & \cite{gl05_nomos} & 2019 &  & \multicolumn{1}{l|}{\cellcolor[HTML]{D9EAD3}{\color[HTML]{D9EAD3} 1}} & \multicolumn{1}{l|}{} & \multicolumn{1}{l|}{} & \multicolumn{1}{l|}{\cellcolor[HTML]{D9EAD3}{\color[HTML]{D9EAD3} 1}} & \multicolumn{1}{l|}{} & \multicolumn{1}{l|}{} & \multicolumn{1}{l|}{\cellcolor[HTML]{D9EAD3}{\color[HTML]{D9EAD3} 1}} & \multicolumn{1}{l|}{} & \multicolumn{1}{l|}{\cellcolor[HTML]{D9EAD3}{\color[HTML]{D9EAD3} 1}} &  \\ \hline
GL06 & \cite{gl06_quicknode} & 2023 &  & \multicolumn{1}{l|}{} & \multicolumn{1}{l|}{\cellcolor[HTML]{D9EAD3}{\color[HTML]{D9EAD3} 1}} & \multicolumn{1}{l|}{} & \multicolumn{1}{l|}{} & \multicolumn{1}{l|}{\cellcolor[HTML]{D9EAD3}{\color[HTML]{D9EAD3} 1}} & \multicolumn{1}{l|}{\cellcolor[HTML]{D9EAD3}{\color[HTML]{D9EAD3} 1}} & \multicolumn{1}{l|}{} & \multicolumn{1}{l|}{} & \multicolumn{1}{l|}{} & \cellcolor[HTML]{D9EAD3}{\color[HTML]{D9EAD3} 1} \\ \hline
GL07 & \cite{gl07_truffle} & n.d. & \multicolumn{1}{c|}{\begin{tabular}[c]{@{}c@{}}New features and trends\end{tabular}} & \multicolumn{1}{l|}{} & \multicolumn{1}{l|}{} & \multicolumn{1}{l|}{} & \multicolumn{1}{l|}{} & \multicolumn{1}{l|}{} & \multicolumn{1}{l|}{} & \multicolumn{1}{l|}{} & \multicolumn{1}{l|}{} & \multicolumn{1}{l|}{} &  \\ \hline
GL08 & \cite{gl08_glaze} & 2023 & \multicolumn{1}{c|}{\begin{tabular}[c]{@{}c@{}}Tool comparison\end{tabular}} & \multicolumn{1}{l|}{} & \multicolumn{1}{l|}{} & \multicolumn{1}{l|}{} & \multicolumn{1}{l|}{} & \multicolumn{1}{l|}{} & \multicolumn{1}{l|}{} & \multicolumn{1}{l|}{} & \multicolumn{1}{l|}{} & \multicolumn{1}{l|}{} &  \\ \hline
GL09 & \cite{gl09_kalos} & 2019 &  & \multicolumn{1}{l|}{\cellcolor[HTML]{D9EAD3}{\color[HTML]{D9EAD3} 1}} & \multicolumn{1}{l|}{} & \multicolumn{1}{l|}{} & \multicolumn{1}{l|}{} & \multicolumn{1}{l|}{} & \multicolumn{1}{l|}{} & \multicolumn{1}{l|}{} & \multicolumn{1}{l|}{} & \multicolumn{1}{l|}{} &  \\ \hline
GL10 & \cite{gl10_spydra} & n.d. &  & \multicolumn{1}{l|}{\cellcolor[HTML]{D9EAD3}{\color[HTML]{D9EAD3} 1}} & \multicolumn{1}{l|}{\cellcolor[HTML]{D9EAD3}{\color[HTML]{D9EAD3} 1}} & \multicolumn{1}{l|}{} & \multicolumn{1}{l|}{\cellcolor[HTML]{D9EAD3}{\color[HTML]{D9EAD3} 1}} & \multicolumn{1}{l|}{} & \multicolumn{1}{l|}{} & \multicolumn{1}{l|}{} & \multicolumn{1}{l|}{\cellcolor[HTML]{D9EAD3}{\color[HTML]{D9EAD3} 1}} & \multicolumn{1}{l|}{} &  \\ \hline
GL11 & \cite{gl11_stratis} & 2023 &  & \multicolumn{1}{l|}{} & \multicolumn{1}{l|}{\cellcolor[HTML]{D9EAD3}{\color[HTML]{D9EAD3} 1}} & \multicolumn{1}{l|}{} & \multicolumn{1}{l|}{\cellcolor[HTML]{D9EAD3}{\color[HTML]{D9EAD3} 1}} & \multicolumn{1}{l|}{} & \multicolumn{1}{l|}{} & \multicolumn{1}{l|}{} & \multicolumn{1}{l|}{} & \multicolumn{1}{l|}{} &  \\ \hline
GL12 & \cite{gl12_stratis} & 2023 &  & \multicolumn{1}{l|}{\cellcolor[HTML]{D9EAD3}{\color[HTML]{D9EAD3} 1}} & \multicolumn{1}{l|}{} & \multicolumn{1}{l|}{} & \multicolumn{1}{l|}{} & \multicolumn{1}{l|}{\cellcolor[HTML]{D9EAD3}{\color[HTML]{D9EAD3} 1}} & \multicolumn{1}{l|}{} & \multicolumn{1}{l|}{\cellcolor[HTML]{D9EAD3}{\color[HTML]{D9EAD3} 1}} & \multicolumn{1}{l|}{} & \multicolumn{1}{l|}{} &  \\ \hline
GL13 & \cite{gl13_azure} & 2019 &  & \multicolumn{1}{l|}{\cellcolor[HTML]{D9EAD3}{\color[HTML]{D9EAD3} 1}} & \multicolumn{1}{l|}{\cellcolor[HTML]{D9EAD3}{\color[HTML]{D9EAD3} 1}} & \multicolumn{1}{l|}{} & \multicolumn{1}{l|}{\cellcolor[HTML]{D9EAD3}{\color[HTML]{D9EAD3} 1}} & \multicolumn{1}{l|}{} & \multicolumn{1}{l|}{} & \multicolumn{1}{l|}{} & \multicolumn{1}{l|}{} & \multicolumn{1}{l|}{} &  \\ \hline
GL14 & \cite{gl14_rahman} & 2018 & \multicolumn{1}{c|}{Personal experiences} & \multicolumn{1}{l|}{} & \multicolumn{1}{l|}{} & \multicolumn{1}{l|}{} & \multicolumn{1}{l|}{} & \multicolumn{1}{l|}{} & \multicolumn{1}{l|}{} & \multicolumn{1}{l|}{} & \multicolumn{1}{l|}{} & \multicolumn{1}{l|}{} &  \\ \hline
GL15 & \cite{gl15_tenderly} & 2022 &  & \multicolumn{1}{l|}{\cellcolor[HTML]{D9EAD3}{\color[HTML]{D9EAD3} 1}} & \multicolumn{1}{l|}{} & \multicolumn{1}{c|}{\cellcolor[HTML]{D9EAD3}{\color[HTML]{D9EAD3} 1}} & \multicolumn{1}{l|}{\cellcolor[HTML]{D9EAD3}{\color[HTML]{D9EAD3} 1}} & \multicolumn{1}{l|}{\cellcolor[HTML]{D9EAD3}{\color[HTML]{D9EAD3} 1}} & \multicolumn{1}{l|}{\cellcolor[HTML]{D9EAD3}{\color[HTML]{D9EAD3} 1}} & \multicolumn{1}{l|}{} & \multicolumn{1}{l|}{} & \multicolumn{1}{l|}{} &  \\ \hline
GL16 & \cite{gl16_polkadot} & 2023 & \multicolumn{1}{c|}{\begin{tabular}[c]{@{}c@{}}New protocols and blockchains\end{tabular}} & \multicolumn{1}{l|}{} & \multicolumn{1}{l|}{} & \multicolumn{1}{l|}{} & \multicolumn{1}{l|}{} & \multicolumn{1}{l|}{} & \multicolumn{1}{l|}{} & \multicolumn{1}{l|}{} & \multicolumn{1}{l|}{} & \multicolumn{1}{l|}{} &  \\ \hline
GL17 & \cite{gl17_comparison} & 2022 & \multicolumn{1}{c|}{\begin{tabular}[c]{@{}c@{}}Tool comparison\end{tabular}} & \multicolumn{1}{l|}{} & \multicolumn{1}{l|}{} & \multicolumn{1}{l|}{} & \multicolumn{1}{l|}{} & \multicolumn{1}{l|}{} & \multicolumn{1}{l|}{} & \multicolumn{1}{l|}{} & \multicolumn{1}{l|}{} & \multicolumn{1}{l|}{} &  \\ \hline
GL18 & \cite{gl18_evmos} & 2023 &  & \multicolumn{1}{l|}{\cellcolor[HTML]{D9EAD3}{\color[HTML]{D9EAD3} 1}} & \multicolumn{1}{l|}{} & \multicolumn{1}{l|}{} & \multicolumn{1}{l|}{} & \multicolumn{1}{l|}{\cellcolor[HTML]{D9EAD3}{\color[HTML]{D9EAD3} 1}} & \multicolumn{1}{l|}{} & \multicolumn{1}{l|}{} & \multicolumn{1}{l|}{} & \multicolumn{1}{l|}{\cellcolor[HTML]{D9EAD3}{\color[HTML]{D9EAD3} 1}} &  \\ \hline
GL19 & \cite{gl19_axelar} & n.d. &  & \multicolumn{1}{l|}{} & \multicolumn{1}{l|}{\cellcolor[HTML]{D9EAD3}{\color[HTML]{D9EAD3} 1}} & \multicolumn{1}{l|}{} & \multicolumn{1}{l|}{} & \multicolumn{1}{l|}{\cellcolor[HTML]{D9EAD3}{\color[HTML]{D9EAD3} 1}} & \multicolumn{1}{l|}{} & \multicolumn{1}{l|}{\cellcolor[HTML]{D9EAD3}{\color[HTML]{D9EAD3} 1}} & \multicolumn{1}{l|}{} & \multicolumn{1}{l|}{\cellcolor[HTML]{D9EAD3}{\color[HTML]{D9EAD3} 1}} &  \\ \hline
GL20 & \cite{gl20_vendia} & 2022 &  & \multicolumn{1}{l|}{} & \multicolumn{1}{l|}{\cellcolor[HTML]{D9EAD3}{\color[HTML]{D9EAD3} 1}} & \multicolumn{1}{l|}{} & \multicolumn{1}{l|}{} & \multicolumn{1}{l|}{} & \multicolumn{1}{l|}{} & \multicolumn{1}{l|}{} & \multicolumn{1}{l|}{} & \multicolumn{1}{l|}{} &  \\ \hline
GL21 & \cite{gl21_fuel} & n.d. &  & \multicolumn{1}{l|}{} & \multicolumn{1}{l|}{\cellcolor[HTML]{D9EAD3}{\color[HTML]{D9EAD3} 1}} & \multicolumn{1}{l|}{} & \multicolumn{1}{l|}{} & \multicolumn{1}{l|}{} & \multicolumn{1}{l|}{\cellcolor[HTML]{D9EAD3}{\color[HTML]{D9EAD3} 1}} & \multicolumn{1}{l|}{\cellcolor[HTML]{D9EAD3}{\color[HTML]{D9EAD3} 1}} & \multicolumn{1}{l|}{} & \multicolumn{1}{l|}{} &  \\ \hline
GL22 & \cite{gl22_microsoft} & 2019 &  & \multicolumn{1}{l|}{\cellcolor[HTML]{D9EAD3}{\color[HTML]{D9EAD3} 1}} & \multicolumn{1}{l|}{} & \multicolumn{1}{l|}{} & \multicolumn{1}{l|}{} & \multicolumn{1}{l|}{} & \multicolumn{1}{l|}{} & \multicolumn{1}{l|}{} & \multicolumn{1}{l|}{} & \multicolumn{1}{l|}{} &  \\ \hline
GL23 & \cite{gl23_scroll} & 2024 &  & \multicolumn{1}{l|}{\cellcolor[HTML]{D9EAD3}{\color[HTML]{D9EAD3} 1}} & \multicolumn{1}{l|}{} & \multicolumn{1}{l|}{} & \multicolumn{1}{l|}{} & \multicolumn{1}{l|}{\cellcolor[HTML]{D9EAD3}{\color[HTML]{D9EAD3} 1}} & \multicolumn{1}{l|}{\cellcolor[HTML]{D9EAD3}{\color[HTML]{D9EAD3} 1}} & \multicolumn{1}{l|}{\cellcolor[HTML]{D9EAD3}{\color[HTML]{D9EAD3} 1}} & \multicolumn{1}{l|}{} & \multicolumn{1}{l|}{} &  \\ \hline
GL24 & \cite{gl24_riady} & 2019 & \multicolumn{1}{c|}{Best practices} & \multicolumn{1}{l|}{} & \multicolumn{1}{l|}{} & \multicolumn{1}{l|}{} & \multicolumn{1}{l|}{} & \multicolumn{1}{l|}{} & \multicolumn{1}{l|}{} & \multicolumn{1}{l|}{} & \multicolumn{1}{l|}{} & \multicolumn{1}{l|}{} &  \\ \hline
GL25 & \cite{gl25_immutable} & n.d. &  & \multicolumn{1}{l|}{\cellcolor[HTML]{D9EAD3}{\color[HTML]{D9EAD3} 1}} & \multicolumn{1}{l|}{} & \multicolumn{1}{c|}{\cellcolor[HTML]{D9EAD3}{\color[HTML]{D9EAD3} 1}} & \multicolumn{1}{l|}{} & \multicolumn{1}{l|}{} & \multicolumn{1}{l|}{} & \multicolumn{1}{l|}{\cellcolor[HTML]{D9EAD3}{\color[HTML]{D9EAD3} 1}} & \multicolumn{1}{l|}{} & \multicolumn{1}{l|}{\cellcolor[HTML]{D9EAD3}{\color[HTML]{D9EAD3} 1}} &  \\ \hline
GL26 & \cite{gl26_txtx} & n.d. &  & \multicolumn{1}{l|}{\cellcolor[HTML]{D9EAD3}{\color[HTML]{D9EAD3} 1}} & \multicolumn{1}{l|}{} & \multicolumn{1}{c|}{\cellcolor[HTML]{D9EAD3}{\color[HTML]{D9EAD3} 1}} & \multicolumn{1}{l|}{} & \multicolumn{1}{l|}{} & \multicolumn{1}{l|}{} & \multicolumn{1}{l|}{\cellcolor[HTML]{D9EAD3}{\color[HTML]{D9EAD3} 1}} & \multicolumn{1}{l|}{\cellcolor[HTML]{D9EAD3}{\color[HTML]{D9EAD3} 1}} & \multicolumn{1}{l|}{} &  \\ \hline
GL27 & \cite{gl27_waves} & 2020 &  & \multicolumn{1}{l|}{\cellcolor[HTML]{D9EAD3}{\color[HTML]{D9EAD3} 1}} & \multicolumn{1}{l|}{\cellcolor[HTML]{D9EAD3}{\color[HTML]{D9EAD3} 1}} & \multicolumn{1}{l|}{} & \multicolumn{1}{l|}{} & \multicolumn{1}{l|}{} & \multicolumn{1}{l|}{} & \multicolumn{1}{l|}{} & \multicolumn{1}{l|}{} & \multicolumn{1}{l|}{} &  \\ \hline
GL28 & \cite{gl28_ignite} & n.d. &  & \multicolumn{1}{l|}{\cellcolor[HTML]{D9EAD3}{\color[HTML]{D9EAD3} 1}} & \multicolumn{1}{l|}{} & \multicolumn{1}{l|}{} & \multicolumn{1}{l|}{} & \multicolumn{1}{l|}{\cellcolor[HTML]{D9EAD3}{\color[HTML]{D9EAD3} 1}} & \multicolumn{1}{l|}{} & \multicolumn{1}{l|}{} & \multicolumn{1}{l|}{} & \multicolumn{1}{l|}{} &  \\ \hline
GL29 & \cite{gl29_comparison} & 2023 & \multicolumn{1}{c|}{\begin{tabular}[c]{@{}c@{}}Tool comparison\end{tabular}} & \multicolumn{1}{l|}{} & \multicolumn{1}{l|}{} & \multicolumn{1}{l|}{} & \multicolumn{1}{l|}{} & \multicolumn{1}{l|}{} & \multicolumn{1}{l|}{} & \multicolumn{1}{l|}{} & \multicolumn{1}{l|}{} & \multicolumn{1}{l|}{} &  \\ \hline
GL30 & \cite{gl30_ligo} & n.d. &  & \multicolumn{1}{l|}{} & \multicolumn{1}{l|}{\cellcolor[HTML]{D9EAD3}{\color[HTML]{D9EAD3} 1}} & \multicolumn{1}{l|}{} & \multicolumn{1}{l|}{\cellcolor[HTML]{D9EAD3}{\color[HTML]{D9EAD3} 1}} & \multicolumn{1}{l|}{} & \multicolumn{1}{l|}{} & \multicolumn{1}{l|}{} & \multicolumn{1}{l|}{} & \multicolumn{1}{l|}{} &  \\ \hline
GL31 & \cite{gl31_polkadot} & 2024 &  & \multicolumn{1}{l|}{} & \multicolumn{1}{l|}{} & \multicolumn{1}{l|}{} & \multicolumn{1}{l|}{} & \multicolumn{1}{l|}{} & \multicolumn{1}{l|}{} & \multicolumn{1}{l|}{} & \multicolumn{1}{l|}{} & \multicolumn{1}{l|}{} &  \\ \hline
GL32 & \cite{gl32_viem} & 2024 &  & \multicolumn{1}{l|}{\cellcolor[HTML]{D9EAD3}{\color[HTML]{D9EAD3} 1}} & \multicolumn{1}{l|}{} & \multicolumn{1}{l|}{} & \multicolumn{1}{l|}{\cellcolor[HTML]{D9EAD3}{\color[HTML]{D9EAD3} 1}} & \multicolumn{1}{l|}{} & \multicolumn{1}{l|}{\cellcolor[HTML]{D9EAD3}{\color[HTML]{D9EAD3} 1}} & \multicolumn{1}{l|}{} & \multicolumn{1}{l|}{\cellcolor[HTML]{D9EAD3}{\color[HTML]{D9EAD3} 1}} & \multicolumn{1}{l|}{} &  \\ \hline
GL33 & \cite{gl33_plutus} & 2021 & \multicolumn{1}{c|}{\begin{tabular}[c]{@{}c@{}}New protocols and blockchains\end{tabular}} & \multicolumn{1}{l|}{} & \multicolumn{1}{l|}{} & \multicolumn{1}{l|}{} & \multicolumn{1}{l|}{} & \multicolumn{1}{l|}{} & \multicolumn{1}{l|}{} & \multicolumn{1}{l|}{} & \multicolumn{1}{l|}{} & \multicolumn{1}{l|}{} &  \\ \hline
GL34 & \cite{gl34_stack} & 2023 &  & \multicolumn{1}{l|}{} & \multicolumn{1}{l|}{\cellcolor[HTML]{D9EAD3}{\color[HTML]{D9EAD3} 1}} & \multicolumn{1}{l|}{} & \multicolumn{1}{l|}{} & \multicolumn{1}{l|}{} & \multicolumn{1}{l|}{} & \multicolumn{1}{l|}{} & \multicolumn{1}{l|}{} & \multicolumn{1}{l|}{} &  \\ \hline
GL35 & \cite{gl35_comparison} & 2023 & \multicolumn{1}{c|}{\begin{tabular}[c]{@{}c@{}}Tool comparison\end{tabular}} & \multicolumn{1}{l|}{} & \multicolumn{1}{l|}{} & \multicolumn{1}{l|}{} & \multicolumn{1}{l|}{} & \multicolumn{1}{l|}{} & \multicolumn{1}{l|}{} & \multicolumn{1}{l|}{} & \multicolumn{1}{l|}{} & \multicolumn{1}{l|}{} &  \\ \hline
GL36 & \cite{gl36_openzepellin} & 2019 &  & \multicolumn{1}{l|}{\cellcolor[HTML]{D9EAD3}{\color[HTML]{D9EAD3} 1}} & \multicolumn{1}{l|}{} & \multicolumn{1}{l|}{} & \multicolumn{1}{l|}{} & \multicolumn{1}{l|}{} & \multicolumn{1}{l|}{} & \multicolumn{1}{l|}{} & \multicolumn{1}{l|}{} & \multicolumn{1}{l|}{} &  \\ \hline
GL37 & \cite{gl37_sam} & 2022 & \multicolumn{1}{c|}{\begin{tabular}[c]{@{}c@{}}Tool comparison\end{tabular}} & \multicolumn{1}{l|}{} & \multicolumn{1}{l|}{} & \multicolumn{1}{l|}{} & \multicolumn{1}{l|}{} & \multicolumn{1}{l|}{} & \multicolumn{1}{l|}{} & \multicolumn{1}{l|}{} & \multicolumn{1}{l|}{} & \multicolumn{1}{l|}{} &  \\ \hline
GL38 & \cite{gl38_presentation} & 2018 & \multicolumn{1}{c|}{Personal experiences} & \multicolumn{1}{l|}{} & \multicolumn{1}{l|}{} & \multicolumn{1}{l|}{} & \multicolumn{1}{l|}{} & \multicolumn{1}{l|}{} & \multicolumn{1}{l|}{} & \multicolumn{1}{l|}{} & \multicolumn{1}{l|}{} & \multicolumn{1}{l|}{} &  \\ \hline
GL39 & \cite{gl39_skale} & 2023 &  & \multicolumn{1}{l|}{\cellcolor[HTML]{D9EAD3}{\color[HTML]{D9EAD3} 1}} & \multicolumn{1}{l|}{} & \multicolumn{1}{l|}{} & \multicolumn{1}{l|}{} & \multicolumn{1}{l|}{} & \multicolumn{1}{l|}{} & \multicolumn{1}{l|}{} & \multicolumn{1}{l|}{} & \multicolumn{1}{l|}{} &  \\ \hline
GL40 & \cite{gl40_cryptoeq} & 2023 & \multicolumn{1}{c|}{\begin{tabular}[c]{@{}c@{}}Tool comparison\end{tabular}} & \multicolumn{1}{l|}{} & \multicolumn{1}{l|}{} & \multicolumn{1}{l|}{} & \multicolumn{1}{l|}{} & \multicolumn{1}{l|}{} & \multicolumn{1}{l|}{} & \multicolumn{1}{l|}{} & \multicolumn{1}{l|}{} & \multicolumn{1}{l|}{} &  \\ \hline
GL41 & \cite{gl41_eosio} & 2019 &  & \multicolumn{1}{l|}{\cellcolor[HTML]{D9EAD3}{\color[HTML]{D9EAD3} 1}} & \multicolumn{1}{l|}{\cellcolor[HTML]{D9EAD3}{\color[HTML]{D9EAD3} 1}} & \multicolumn{1}{l|}{} & \multicolumn{1}{l|}{\cellcolor[HTML]{D9EAD3}{\color[HTML]{D9EAD3} 1}} & \multicolumn{1}{l|}{} & \multicolumn{1}{l|}{} & \multicolumn{1}{l|}{} & \multicolumn{1}{l|}{} & \multicolumn{1}{l|}{} &  \\ \hline
GL42 & \cite{gl42_near} & 2023 & \multicolumn{1}{c|}{\begin{tabular}[c]{@{}c@{}}Resource repository\end{tabular}} & \multicolumn{1}{l|}{} & \multicolumn{1}{l|}{} & \multicolumn{1}{l|}{} & \multicolumn{1}{l|}{} & \multicolumn{1}{l|}{} & \multicolumn{1}{l|}{} & \multicolumn{1}{l|}{} & \multicolumn{1}{l|}{} & \multicolumn{1}{l|}{} &  \\ \hline
GL43 & \cite{gl43_aiken} & 2023 &  & \multicolumn{1}{l|}{} & \multicolumn{1}{l|}{\cellcolor[HTML]{D9EAD3}{\color[HTML]{D9EAD3} 1}} & \multicolumn{1}{l|}{} & \multicolumn{1}{l|}{\cellcolor[HTML]{D9EAD3}{\color[HTML]{D9EAD3} 1}} & \multicolumn{1}{l|}{} & \multicolumn{1}{l|}{} & \multicolumn{1}{l|}{} & \multicolumn{1}{l|}{\cellcolor[HTML]{D9EAD3}{\color[HTML]{D9EAD3} 1}} & \multicolumn{1}{l|}{} &  \\ \hline
GL44 & \cite{gl44_zycrypto} & 2021 &  & \multicolumn{1}{l|}{\cellcolor[HTML]{D9EAD3}{\color[HTML]{D9EAD3} 1}} & \multicolumn{1}{l|}{} & \multicolumn{1}{l|}{} & \multicolumn{1}{l|}{} & \multicolumn{1}{l|}{\cellcolor[HTML]{D9EAD3}{\color[HTML]{D9EAD3} 1}} & \multicolumn{1}{l|}{\cellcolor[HTML]{D9EAD3}{\color[HTML]{D9EAD3} 1}} & \multicolumn{1}{l|}{} & \multicolumn{1}{l|}{} & \multicolumn{1}{l|}{} &  \\ \hline
GL45 & \cite{gl45_openzeppelin_errors} & 2023 & \multicolumn{1}{c|}{\begin{tabular}[c]{@{}c@{}}New features and trends\end{tabular}} & \multicolumn{1}{l|}{} & \multicolumn{1}{l|}{} & \multicolumn{1}{l|}{} & \multicolumn{1}{l|}{} & \multicolumn{1}{l|}{} & \multicolumn{1}{l|}{} & \multicolumn{1}{l|}{} & \multicolumn{1}{l|}{} & \multicolumn{1}{l|}{} &  \\ \hline
GL46 & \cite{gl46_pomposi} & 2021 &  & \multicolumn{1}{l|}{} & \multicolumn{1}{l|}{} & \multicolumn{1}{l|}{} & \multicolumn{1}{l|}{\cellcolor[HTML]{D9EAD3}{\color[HTML]{D9EAD3} 1}} & \multicolumn{1}{l|}{} & \multicolumn{1}{l|}{} & \multicolumn{1}{l|}{} & \multicolumn{1}{l|}{} & \multicolumn{1}{l|}{} &  \\ \hline
GL47 & \cite{gl47_solidity_survey} & 2023 & \multicolumn{1}{c|}{\begin{tabular}[c]{@{}c@{}}Empirical results or surveys\end{tabular}} & \multicolumn{1}{l|}{} & \multicolumn{1}{l|}{} & \multicolumn{1}{l|}{} & \multicolumn{1}{l|}{} & \multicolumn{1}{l|}{} & \multicolumn{1}{l|}{} & \multicolumn{1}{l|}{} & \multicolumn{1}{l|}{} & \multicolumn{1}{l|}{} &  \\ \hline
GL48 & \cite{gl48_infura} & n.d. &  & \multicolumn{1}{l|}{\cellcolor[HTML]{D9EAD3}{\color[HTML]{D9EAD3} 1}} & \multicolumn{1}{l|}{} & \multicolumn{1}{l|}{} & \multicolumn{1}{l|}{} & \multicolumn{1}{l|}{\cellcolor[HTML]{D9EAD3}{\color[HTML]{D9EAD3} 1}} & \multicolumn{1}{l|}{\cellcolor[HTML]{D9EAD3}{\color[HTML]{D9EAD3} 1}} & \multicolumn{1}{l|}{} & \multicolumn{1}{l|}{} & \multicolumn{1}{l|}{} & \multicolumn{1}{r|}{\cellcolor[HTML]{D9EAD3}{\color[HTML]{D9EAD3} 1}} \\ \hline
GL49 & \cite{gl49_thirdweb} & n.d. &  & \multicolumn{1}{l|}{\cellcolor[HTML]{D9EAD3}{\color[HTML]{D9EAD3} 1}} & \multicolumn{1}{l|}{} & \multicolumn{1}{l|}{} & \multicolumn{1}{l|}{} & \multicolumn{1}{l|}{\cellcolor[HTML]{D9EAD3}{\color[HTML]{D9EAD3} 1}} & \multicolumn{1}{l|}{\cellcolor[HTML]{D9EAD3}{\color[HTML]{D9EAD3} 1}} & \multicolumn{1}{l|}{\cellcolor[HTML]{D9EAD3}{\color[HTML]{D9EAD3} 1}} & \multicolumn{1}{l|}{} & \multicolumn{1}{l|}{} &  \\ \hline
GL50 & \cite{gl50_zksync} & 2024 &  & \multicolumn{1}{l|}{\cellcolor[HTML]{D9EAD3}{\color[HTML]{D9EAD3} 1}} & \multicolumn{1}{l|}{} & \multicolumn{1}{l|}{} & \multicolumn{1}{l|}{} & \multicolumn{1}{l|}{} & \multicolumn{1}{l|}{} & \multicolumn{1}{l|}{\cellcolor[HTML]{D9EAD3}{\color[HTML]{D9EAD3} 1}} & \multicolumn{1}{l|}{} & \multicolumn{1}{l|}{\cellcolor[HTML]{D9EAD3}{\color[HTML]{D9EAD3} 1}} &  \\ \hline
GL51 & \cite{gl51_besu} & 2018 &  & \multicolumn{1}{l|}{\cellcolor[HTML]{D9EAD3}{\color[HTML]{D9EAD3} 1}} & \multicolumn{1}{l|}{} & \multicolumn{1}{c|}{\cellcolor[HTML]{D9EAD3}{\color[HTML]{D9EAD3} 1}} & \multicolumn{1}{l|}{} & \multicolumn{1}{l|}{\cellcolor[HTML]{D9EAD3}{\color[HTML]{D9EAD3} 1}} & \multicolumn{1}{l|}{} & \multicolumn{1}{l|}{\cellcolor[HTML]{D9EAD3}{\color[HTML]{D9EAD3} 1}} & \multicolumn{1}{l|}{} & \multicolumn{1}{l|}{} &  \\ \hline
GL52 & \cite{gl52_algorand} & 2023 &  & \multicolumn{1}{l|}{\cellcolor[HTML]{D9EAD3}{\color[HTML]{D9EAD3} 1}} & \multicolumn{1}{l|}{} & \multicolumn{1}{l|}{} & \multicolumn{1}{l|}{\cellcolor[HTML]{D9EAD3}{\color[HTML]{D9EAD3} 1}} & \multicolumn{1}{l|}{} & \multicolumn{1}{l|}{} & \multicolumn{1}{l|}{} & \multicolumn{1}{l|}{} & \multicolumn{1}{l|}{} &  \\ \hline
GL53 & \cite{gl53_kaleido} & n.d. &  & \multicolumn{1}{l|}{\cellcolor[HTML]{D9EAD3}{\color[HTML]{D9EAD3} 1}} & \multicolumn{1}{l|}{} & \multicolumn{1}{l|}{} & \multicolumn{1}{l|}{} & \multicolumn{1}{l|}{\cellcolor[HTML]{D9EAD3}{\color[HTML]{D9EAD3} 1}} & \multicolumn{1}{l|}{\cellcolor[HTML]{D9EAD3}{\color[HTML]{D9EAD3} 1}} & \multicolumn{1}{l|}{\cellcolor[HTML]{D9EAD3}{\color[HTML]{D9EAD3} 1}} & \multicolumn{1}{l|}{} & \multicolumn{1}{l|}{\cellcolor[HTML]{D9EAD3}{\color[HTML]{D9EAD3} 1}} &  \\ \hline
GL54 & \cite{gl54_alchemy} & n.d. &  & \multicolumn{1}{l|}{\cellcolor[HTML]{D9EAD3}{\color[HTML]{D9EAD3} 1}} & \multicolumn{1}{l|}{\cellcolor[HTML]{D9EAD3}{\color[HTML]{D9EAD3} 1}} & \multicolumn{1}{c|}{\cellcolor[HTML]{D9EAD3}{\color[HTML]{D9EAD3} 1}} & \multicolumn{1}{l|}{} & \multicolumn{1}{c|}{\cellcolor[HTML]{D9EAD3}{\color[HTML]{D9EAD3} 1}} & \multicolumn{1}{l|}{\cellcolor[HTML]{D9EAD3}{\color[HTML]{D9EAD3} 1}} & \multicolumn{1}{l|}{\cellcolor[HTML]{D9EAD3}{\color[HTML]{D9EAD3} 1}} & \multicolumn{1}{l|}{} & \multicolumn{1}{l|}{\cellcolor[HTML]{D9EAD3}{\color[HTML]{D9EAD3} 1}} &  \\ \hline
GL55 & \cite{gl55_moralis} & n.d. &  & \multicolumn{1}{l|}{\cellcolor[HTML]{D9EAD3}{\color[HTML]{D9EAD3} 1}} & \multicolumn{1}{l|}{} & \multicolumn{1}{c|}{\cellcolor[HTML]{D9EAD3}{\color[HTML]{D9EAD3} 1}} & \multicolumn{1}{l|}{} & \multicolumn{1}{l|}{\cellcolor[HTML]{D9EAD3}{\color[HTML]{D9EAD3} 1}} & \multicolumn{1}{l|}{\cellcolor[HTML]{D9EAD3}{\color[HTML]{D9EAD3} 1}} & \multicolumn{1}{l|}{\cellcolor[HTML]{D9EAD3}{\color[HTML]{D9EAD3} 1}} & \multicolumn{1}{l|}{} & \multicolumn{1}{l|}{\cellcolor[HTML]{D9EAD3}{\color[HTML]{D9EAD3} 1}} &  \\ \hline
\end{tabular}}
\end{table}

\subsubsection{Tool, platform/service, language, method/technique, model, process, or framework}

\begin{figure*}[htbp]
\centerline{\includegraphics[scale=0.7]{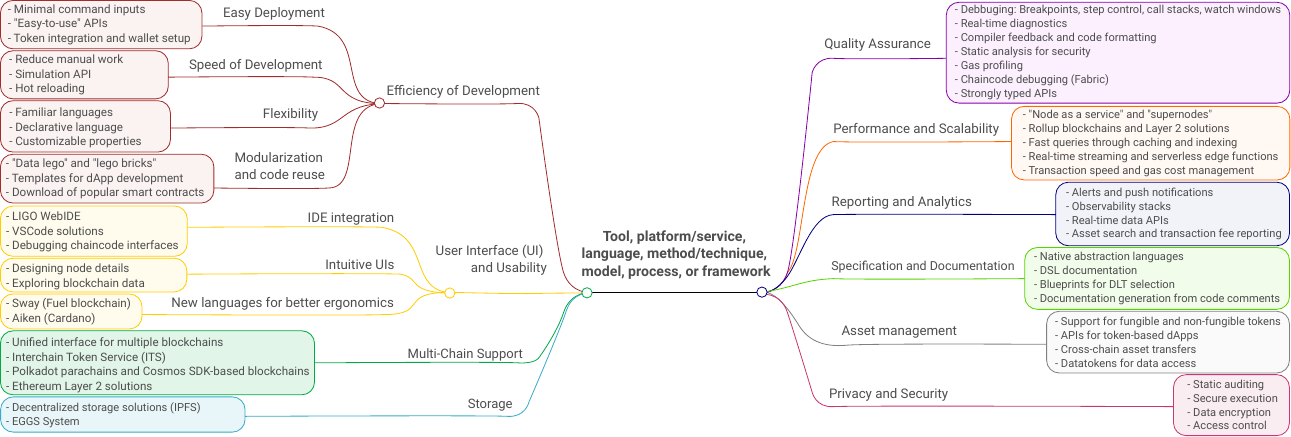}}
\caption{Overview of sources related to Tool, platform/service, language, method/technique, model, process, or frameworks.}
\label{fig:group1}
\end{figure*}

In this analysis, we identified features impacting \textbf{efficiency of development}, including easy deployment, speed, flexibility, and modularization. Some platforms have proposed \textit{easy deployment} with minimal command inputs [GL09, GL28, GL54], offering ``easy-to-use" APIs [GL48, GL25] for blockchain development, with features such as token integration and wallet setup [GL54, GL55]. 

Platforms such as Azure cover the entire development cycle [GL13, GL27, GL41, GL49] with encapsulated tools [WL05]. [GL52] emphasizes the importance of on-premises networks in providing a sandbox environment before mainnet deployment. Blockchain as a service also simplifies network deployment with robust cloud tools. With simple push-buttons, key management, node backups, log streaming, and single sign-on to any blockchain-based business applications can be performed in the cloud and region of the user's choice [GL53].

These tools enhance \textit{speed of development} and ``save time'' by reducing manual work [GL15, GL53]. Using a simulation API, for example, allows users to visualize transaction results and submit them more confidently [GL15, GL41]. Hot reloading is another feature that automatically reloads the source code of a web app and the blockchain when changes are made without setting up multiple steps before compiling the smart contract [GL28, GL36]. Furthermore, these tools can help developers of all experience levels and promote proper integration between Web2 and Web3 components [GL52].

In addition to APIs, third-party libraries, SDKs, or services are essential resources to abstract the complexities of using blockchain platforms and integrate existing systems~[WL04, GL36, GL39, GL25, GL28]. Moreover, new and challenging concepts, such as Parachains in Polkadot, can be simplified through SDKs. [WL04] includes handlers for multiple pallets responsible for Cross-Consensus Message Format (XCM), Asset, and Horizontal Relay-routed Message Passing (HRMP) calls using customized parameters. These parameters include the source and destination Parachains, the token, the amount being transferred, and the recipient's address. Both Polkadot and Cosmos ecosystems present development challenges. However, the community enables solutions to provide seamless communication with other chains, apps with a single deployment and features that automatically create a web app to interact with the blockchain, including CosmJS [GL18, GL28].

Platforms' increasingly common automation processes often influence the ease of deployment and speed of development. For instance, [WL06] clone source code, orchestrate verification, and serve source code over a REST API. [GL15] provides `Web3 Actions' to automate workflows and respond to on-chain and off-chain events in less than a second.

Solutions have also been focused on providing \textit{flexibility} to support better developers transitioning from Web2 to Web3. In this regard, platforms have made efforts to allow developers to ``build with your usual dev tools" [GL23], develop custom extensions in familiar languages (such as JavaScript, Go, Rust, or C), and code workflows to interact with smart contracts using declarative language [GL26], for example. Furthermore, customizing properties (consensus mechanisms, blockchain type, root configurations) enables more flexibility in developing dApps. This feature makes it easier for developers to onboard the blockchain domain, as they can build dApps using their preferred stack. [GL04] provides a simple and extensible way to modify existing data models and add new ones with minimal changes to the source code required by the developer.

Another type of resource platform present is \textit{modularization and code reuse} contributing to development efficiency. The Ocean protocol [WL05] serves as ``data lego" to simplify development through ``composability" and data tokens to connect data assets with Decentralized Finance (DeFi) tools. [GL27] also introduces the concept of ``lego bricks", which provide reusable code snippets for easier development. [GL52] offers a set of templates to use as a starting point for dApp development. In addition, [GL22] allows developers to discover and download popular smart contracts from OpenZeppelin, one of the developers' most widely used libraries. In this sense, its reputation and audited smart contracts can influence the DEx. In terms of APIs, we also found modular and composable APIs [GL32] and API infrastructure that allows developers to leverage pre-built modules [GL44]. Moreover, [WL01] introduces the concept of ``design pattern as a service". This solution provides data management services, smart contract design services, and auxiliary services based on design patterns to better leverage the unique properties of blockchain and help developers save time using modular and pre-built services.

Regarding \textbf{User Interface (UI) and usability}, the results mainly highlight the integration of tools with Integrated Development Environments (IDEs). LIGO WebIDE [GL30] facilitates the creation of LIGO projects and supports writing and testing smart contracts in CameLIGO and JsLIGO with syntax highlighting. Visual Studio Code (VSCode) has been widely used with solutions such as [GL13], providing a UI that allows developers to interact with their contracts within the IDE. Meanwhile, [GL10] offers a plugin with a powerful, intuitive interface for debugging chain codes without requiring complex setup procedures. Waves IDE [GL27] generates interfaces for smart contracts and enables running transfer and exchange transactions or simply writing data to the blockchain. 

Using graphical interfaces, [GL20] allows developers to design node details, auto-populate the query arguments input box based on the provided smart contract query, and edit smart contracts using a formatted editor instead of a plain text box. With [GL41], EOS developers can effortlessly explore blocks within their development nodes and inspect smart contract account details directly on the blockchain. Uploading smart contracts is streamlined with a user-friendly drag-and-drop interface. Moreover, [GL11] provides visual feedback similar to IntelliSense and presents code maps and visual diagrams for smart contracts. [GL34] presents a UI that shows how to interact with an EVM-compatible blockchain and its smart contracts. This solution has addressed challenges by combining the server-rendering paradigm introduced by React and Next.js with the client-side nature of user-facing dApps.

Other solutions are more specific, such as white-label RPC services [GL06], an in-browser request sandbox [GL54], a repository that serves as a universal UI for experimenting with cross-chain transactions [WL04], and a no-code portal for token management [GL19]. In addition, platforms have introduced new languages based on Rust to address ergonomics challenges. For example, Sway [GL21] (Fuel blockchain) uses a domain-specific language (DSL), and Aiken [GL43] (Cardano) offers server capabilities with editor integrations.

The platforms also offer \textbf{multi-chain support} as key features to attract developers from different ecosystems [GL04, GL12, GL19, GL44, GL54, GL06, GL49]. [GL19] introduces the `Interchain Token Service' (ITS), designed to scale token operations across multiple chains, enhancing interoperability. This service aims to support new and existing tokens, preserving their fungibility and native functionalities on EVM-compatible chains. [GL55] facilitates cross-chain token transfers through accessible APIs, including DeFi services.

Furthermore, [WL04] offers direct operations to query and manage assets on parachains on Polkadot, with an architecture that facilitates communication between parachains and other blockchains. This solution includes a map of XCM pallets (modules that simplify cross-chain communication) provided by each compatible para chain, allowing users to query these data as needed. Cosmos also faces the challenge of bridging the gap between its internal ecosystem and other Web3 ecosystems. [GL18] introduce EVM extensions that enable seamless communication with other chains through a single deployment. In [GL28], developers can launch a simple Cosmos SDK-based faster, providing interoperability by supporting the Inter-Blockchain Communication (IBC) protocol.

Scroll [GL23], a Layer 2 network described as ``built by Ethereum devs for Ethereum devs'', provides interoperability bridging ETH to different networks. It aims to deliver an accessible scaling solution while preserving Ethereum's core features and adding new capabilities, such as zero knowledge. [GL51] presents the integration of Monax's Bos deployment tool into Burrow Deploy to support Ethereum smart contract systems, making the deployment and management of Ethereum smart contracts in permissioned environments possible. Moreover, [GL53] presents solutions for permissioned ecosystems by supporting private blockchains and consortium architectures through a consortia management platform.

Related to the category of \textbf{quality assurance}, \textit{debugging} was one of the most frequently discussed themes [GL10, GL13, GL15, GL30, GL43, GL46, GL10]. [GL13] integrated Truffle Debugger into VSCode, providing debugging features (breakpoints, step control, call stacks, watch windows, and Intellisense pop-ups). [GL30] provides a VSCode extension that integrates with the Debbug Adapter Protocol (DAP) with an intuitive UI. In addition, some solutions offer real-time diagnostics with parser and typer error recovery [GL30], useful compiler feedback and automatic code formatting [GL43], bi-directional type checker with precise error message [GL05], and static analysis to detect security issues [GL11]. [WL06] provides a standalone library for verifying code accuracy using a Rust language source-to-bytecode verification routine. On the other hand, tools such as [GL10] designed features for debugging chain code in networks using Hyperledger Fabric.

We also found solutions that present a 'Gas Profiler' to develop gas-optimized smart contracts [GL15]. [GL32] proposes strongly typed APIs, type inference, and static validation. Finally, [WL02] presents two main features related to analyzing Bitcoin APIs: i) applications for violations of API heuristics and guidelines and ii) identifying the best practices for addressing problematic API issues.

Regarding \textbf{performance and scalability}, solutions offer `node as a service' and `supernodes' for read-heavy workloads [GL15, GL48, GL54, GL53], along with optimized bundle sizes and enhanced task execution [GL32]. Platforms provide rollup blockchains [GL06] and Layer 2 solutions with EVM compatibility [GL23], including parallel transaction processing through UTXO model [GL21]. Related to fast queries, [GL44] introduces available API nodes that cache and index blockchain data, achieving results 10 times faster, along with APIs-core that offer extremely low throughput. Meanwhile, [GL06] enables real-time data streaming and serverless edge functions for blockchain data to improve performance, and [GL54] uses subgraphs through custom GraphQL APIs. Moreover, this platform aims to enhance performance by improving transaction speed and managing gas costs with a `Gas Manager API', sending batch transactions using a reliable `Bundler API', and completing transaction history in one call with `Transfer API' [GL49] allows developers to mint tokens and perform on-chain actions with automatic nonce management, transaction queuing, and gas-optimized retries, which promises to autoscale dApps. [WL01] introduces a hybrid on-chain and off-chain service for storing critical and immutable data on-chain while keeping other data off-chain to enhance data reading efficiency. This service lets users define the data schema and specify which attributes are stored on-chain.

\begin{figure*}[h]
\centerline{\includegraphics[scale=0.7]{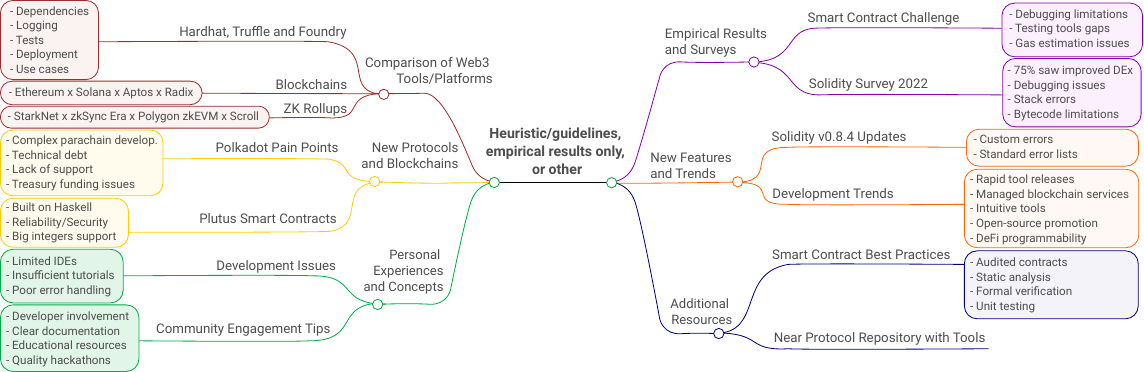}}
\caption{Overview of sources related to heuristic/guidelines, empirical results only, or other.}
\label{fig:group2}
\end{figure*}

In the \textbf{reporting and analytics}, we found features related to alerts and push notifications, such as instant on-chain alerts~[GL06] and fast push notifications through webhooks~[GL54]. Additionally, solutions offer observability stacks with reporting, analytics, and monitoring capabilities [GL15, GL54]. The solutions enhance real-time data capabilities with APIs that stream live general on-chain data to backends and offer self-serve query functionalities to support these functions. These data are related to blocks, transactions, logs, raw and decoded data, transaction labeling, real-time crypto prices [GL55], assets, owners, metadata [GL25], and transaction fee reports [GL26]. These queries are usually enabled by services that manage smart contract events. For instance, [GL51] introduces an ETL/execution event service, and [GL54] uses custom webhooks for custom events.

Regarding \textbf{specification and documentation}, [GL02] aims to enhance the DEx by creating a native abstraction language that allows developers to represent their specifications as a programming language to interact with blockchains. In turn, [GL26] provides documentation for domain-specific languages (DSL) using a native syntax proposed by the solution, enabling developers to define which blockchains and networks to use, what data to retrieve, and which transactions to broadcast, as well as the dependencies between resources. [GL04] proposes a framework for developers to define a blueprint for selecting a new DLT for InteroperaChain. This blueprint includes a decision tree to guide developers on data storage capabilities and the availability of a client library written in a language supported by the solution. Finally, [GL43] allows documentation to be generated directly from code comments.

Several features related to \textbf{asset management} support different token types, from fungible to non-fungible tokens [GL05, GL25, GL54, GL55, GL53]. Some provide APIs to enable transfers, real-time pricing, and custom ecosystems. For example, [GL53] offers a sandbox to simplify the development of Central Bank Digital Currencies (CBDCs) and provides full-stack tools for designing, building, and testing digital currencies. [GL18] and [GL19] allow asset transfers across different chains, ensuring interoperability. [WL04] enables the mapping of assets compatible with each Parachain on Polkadot. On the other hand, [WL05] serves as an on-ramp through data tokens (fungible ERC20 tokens to access specific data services in crypto ecosystems).

In the context of \textbf{privacy and security}, features aim to protect data integrity, manage permissions, and ensure secure transactions. The authorities and embedded permission service [WL01] provide access control, while data encryption and hash integrity protect sensitive information through encryption and integrity checking. Additionally, the dynamic binding service supports initially unknown participant addresses. Compute-to-Data [WL05] provides a way to share or monetize one’s data while preserving privacy. Some platforms also provide mechanisms for the secure execution of smart contracts. For example, off-chain computations that perform checks on smart contracts [GL12] ensure the correctness of computations without exposing critical data on-chain. In addition, [GL19] provides battle-tested proof-of-stake verification.

In terms of development security, [WL27] brings the concept of static auditing to smart contracts. Additionally, solutions aimed at projects with Ethereum-level security [GL25] and platforms like [GL26] simplify using secure multi-signature flows and secure enclaves in the cloud, abstracting complexity through runbooks. Furthermore, The built-in key signing daemon [GL51] helps manage secure key signing operations, and account contracts [GL54] and wallet APIs [GL55] provide secure frameworks for managing Web3 assets and wallets. Digital asset platforms [GL49] provide robust tools for managing and securing assets, while user authentication and integration services [GL53] facilitate connecting to Web3.

In terms of \textbf{storage}, [GL03] presents the External Global Grouped Storage (EGGS) system, which enables the platform to manage large volumes of data by offering developers easy access to smart contracts. To draw an analogy with the web2 environment, developers would open a file and write to it, and their application would seamlessly access the storage. [GL48, GL06] provide Interplanetary File System (IPFS) gateways and pinning services for decentralized storage solutions.

\subsubsection{Heuristic/guidelines, empirical results only, or other}

\begin{figure}[h]
\centerline{\includegraphics[scale=0.4]{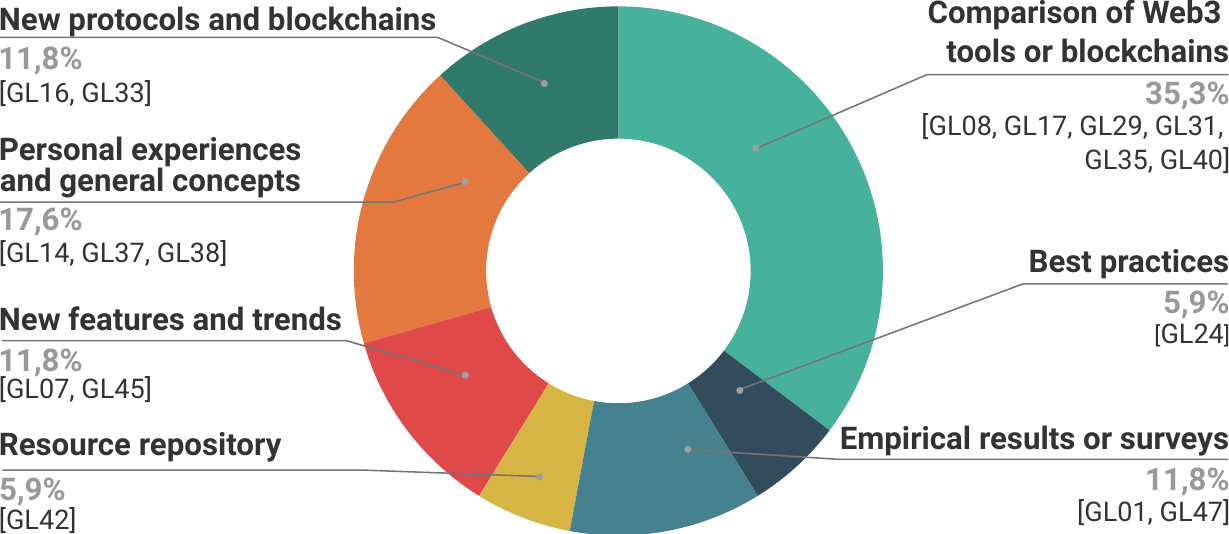}}
\caption{Main categories regarding the heuristic/guidelines, empirical results only, or other sources.}
\label{fig:guidelines}
\end{figure}

Figure~\ref{fig:guidelines} shows the categories addressed within sources. Most sources in this category presented \textbf{comparison of Web3 tools or platforms}. [GL17] compared and presented practices using dependencies, logging, tests, and deployment between Foundry and Hardhat. Similarly, [GL29] compared Truffle and Hardhat on the Celo Blockchain, emphasizing core features, use cases, key differences, and factors to consider. Finally, [GL35] compared Hardhat, Truffle, and Foundry, emphasizing distinctive characteristics, pros, cons, and the best fit for each, as well as key commands and practical scenarios. Meanwhile, [GL40] thoroughly compared the Ethereum, Solana, Aptos, and Radix blockchains. [GL31] provided information on Polkadot and, at the end, offered a brief comparison of parachains and smart contracts, considering aspects such as speed of development, ease of deployment, complexity of logic, maintenance overhead, level of customization, strict resource control, native chain features, and scalability. On a more specific field, [GL08] conducted an in-depth analysis of DEx in ZK Rollups, such as StarkNet, zkSync Era, Polygon zkEVM, and Scroll, which are different and detailed aspects of the development lifecycle.

Q/A forums have also discussed the DEx of \textbf{new protocols and blockchains}. In [GL16], the thread titled ``Developer experience must be our \#1 priority" in the Polkadot forum highlights the major pain points development teams face in Polkadot, acknowledging that ``parachain development is rough". According to the developers, specific technologies used in parachains (such as Substrate) are highly complex and carry substantial technical debt. Moreover, parachain teams feel unsupported and unheard by the creators. Financial issues were also discussed, including uncertainty around Treasury funding. In turn, developers discuss the DEx of Plutus smart contracts in [GL33]. They discuss the language's benefits, such as its reliability and security due to being built in the Haskell language. It supports big integers, simplifies arithmetic, and leverages existing Haskell tools and libraries. Optimizing blockchain storage space was also emphasized.

[GL14] and [GL38] are presentations by developers discussing \textbf{personal experiences and general concepts} in blockchain. [GL38] identifies issues such as needing more IDEs, better education/tutorials, community error support, and improved error message handling in smart contract development platforms. In turn, [GL14] highlights personal experiences in enterprise blockchain development. [GL37] host suggests different tips for engaging the Web3 community, such as involving developers to make technologies successful, having clear documentation, support, and educational resources, balancing technical complexity not to alienate beginners, and providing high-quality hackathons.

Considering sources that present \textbf{empirical results or surveys}, [GL01] presented findings regarding challenges in DExs with smart contracts. At the time of the study, developers participating in the experiments reported that debugging and testing were hindered by the lack of practical tools to provide technical assistance at the source code level, as well as the absence of gas usage estimation and efficient tools for unit and integration testing, in addition to the inadequate interface for viewing state variables. A survey conducted by the Solidity community [GL47] in 2022 also supports the aforementioned issues. According to the survey, 0.9\% of those who felt that the Solidity experience had worsened reported frequent debugging issues, followed by deep stack errors and bytecode size limitations. However, most developers surveyed (+75\%) believe the Solidity DEx has improved over the past year. 

Regarding \textbf{new features and trends}, to address error handling issues, Solidity v0.8.4 introduced the ability to define custom errors, as shown by [GL45]. This source explains that OpenZeppelin proposed a list of standard errors for common token types (ERC-20, ERC-721, and ERC-1155). It also presents results on using custom errors to reduce costs. [GL07] briefly presented five trends that impact blockchain DEx: i) the speed of release of development tools; ii) managed blockchain services; iii) evolving DEx with intuitive tools; iv) corporations promoting open-source software; and v) increasing programmability across industries, led by DeFi.

Other sources were more specific, such as [GL24], which provided a list of \textbf{best practices} for smart contract development. These practices include using audited contracts, prioritizing DEx, employing static analysis tools, considering formal verification, writing unit tests, and more. Finally, [GL42] made available an extensive \textbf{repository of resources} with tools for improving DEx using the Near Protocol.

\begin{tcolorbox}[right=0.1cm,left=0.1cm,top=0.1cm,bottom=0.1cm]
\small\textbf{Answer to RQ2:} \textit{Our analysis shows that 74.19\% of sources (46) fall into the first group, covering tools, platforms/services, languages, methods/techniques, models, processes, or frameworks. The remaining 25.81\% (16 sources) comprise the second group, which includes resources such as guidelines, empirical findings, and other insights. Within the first group, the most frequently mentioned aspects were development efficiency, multi-network support, and UI/usability. On the other hand, the most discussed topics in the second group included comparisons of Web3 tools, personal experiences, general concepts, and the complexities of protocols and blockchains.}
\end{tcolorbox}

\subsection{RQ3) In what ways have the sources discussed in the literature been shaping the BcDEx in practice?}
\label{sec:rq3}
In this section, we discuss how the sources have been shaping the BcDEx, focusing on their effects and benefits.

\subsubsection{Simplifying complexity through abstraction and enhancing usability}
The abstraction of complexity and heterogeneity of blockchain connectors for developers has been one of the primary focuses of blockchain solutions identified [GL04]. Solutions such as [GL02], for instance, abstract the need for protocol design and allow developers to interact with arbitrary blockchains through simple programmatic syntax. The proposed platforms enable developers to avoid infrastructure setup and maintenance [GL15, GL10], allowing them to concentrate on key business logic while abstracting trivial details [GL09]. Features such as interfaces [GL20], map interaction [GL05], and transaction simulation [GL15] can make the development process easier and more efficient and provide developers with an ``invisible blockchain experience'' [GL39].

\subsubsection{Facilitating the adoption of blockchain technologies}
Such solutions can ensure the frictionless adoption of new technologies and reduce the entry barriers for developers in blockchain technology [WL03, GL28]. This issue is essential for promoting its use and enabling a smooth transition from Web2 to Web3 technologies [GL03, GL08, GL44, GL52]. Most of the presented solutions aim to provide a user-friendly experience for developers of all skill levels, making it easier for those with little Web3 experience to get started [GL10]. For example, solution [WL07] promises that technical and non-technical developers can create contracts more efficiently. Another relevant point is that platforms have introduced features to support different ecosystems and allow developers to work with specific Web2 stacks, such as [GL11], tailored for C\# developers. Similarly, solutions such as [GL22] empower Ethereum developers, both public and private networks.

\subsubsection{Impact on developer productivity}
Reducing programmer overhead [GL05] and saving development time and operational costs [WL01, WL04, GL43, GL25] can positively impact developer productivity, as expressed in [WL06, GL08, GL10, GL22]. Several solutions offer multiple services on a single platform, such as [WL05, GL15, GL19, GL23, GL25, GL48, GL54, GL55, GL06, GL53, GL49]. These resources enable developers to access services for most phases of decentralized application development, from libraries that abstract the complexity of blockchain connectivity and functionalities to infrastructures of varying sizes, including the complete setup of nodes and networks. These tools also increase development speed, allowing teams to work together more efficiently without relying on isolated tools, thus improving project flow and overall productivity.

\subsubsection{Education, training, and support for blockchain developers}
This support is a key factor in the BcDEx. Resources such as guidelines and heuristics help developers familiarize themselves with key blockchain concepts [WL02, GL01]. These educational resources can provide insights into how to write efficient and maintainable code [GL24, GL27]. Platforms such as [GL42] contribute by making available a wide range of resources to support their projects, and programs such as hackathons and boot camps contribute to hands-on training [GL52]. Furthermore, community engagement and collaboration are equally important. Resources and guidelines based on presenting and sharing real-world experiences allow developers to learn from practical challenges faced by others [GL14, GL16, GL33, GL38, GL47]. Moreover, developers benefit from feedback loops within the blockchain community, which can be instrumental in selecting the best tools and frameworks for specific projects through knowledge sharing and updates on new features [GL17, GL29, GL35, GL40].

\subsubsection{BcDEx evaluation} In the Polkadot ecosystem, researchers evaluated DEx by comparing how experienced developers performed using different interfaces: the PolkadotJS XCM UI versus the ParaSpell XCM SDK [WL04]. The results demonstrated improved usability aspects with the ParaSpell SDK, as developers completed tasks faster and with fewer errors. In the Polkadot Q\&A forums~[GL16], the authors emphasized the importance of developer engagement and deployment of new applications as indicators of success, as previous efforts effectively attract and support new developers.

[GL01] assessed the challenges faced by smart contract developers through a task analysis. This approach provided reports on how developers interact with leading smart contract development tools. Finally, [WL06] uses the Sui Blockchain and Move Language approach to verify and display smart contract source code. The study aims to assess how often developers consult source code, and plans to incorporate quality indicators to highlight software quality signals.

\begin{tcolorbox}[right=0.1cm,left=0.1cm,top=0.1cm,bottom=0.1cm]
\small\textbf{Answer to RQ3:} \textit{We identified five key perspectives through which sources in the literature have been shaping BcDEx in practice: abstraction for usability, blockchain adoption facilitation, productivity impact, developer education and support, and BcDEx evaluation. Together, these aspects enhance BcDEx by lowering entry barriers, optimizing workflows, and supporting continuous improvement.}
\end{tcolorbox}

\section{Discussion}
\label{sec:discussion}

Our study provides a detailed examination of BcDEx, analyzing the distribution of literature, practical implementations, and their impact on BcDEx. In the following sections, we discuss the primary takeaways in relation to each RQ.

\subsection{In RQ1, we asked about the distribution and nature of DEx in blockchain solutions}
The bibliographic analysis reveals a distinct evolution in discussions on BcDEx, primarily fueled by industry sources~[GL52, GL54]. While contributions from academic sources remain limited, the steady growth in industry publications, especially blogs and company websites, indicates a strong response to the need for improved BcDEx. This emphasis reflects a practical demand for developer-oriented solutions as blockchain complexity increases. In addition, the increase in both academic and industry publications in 2023 reflects a rising awareness of DEx as an important factor in sustaining blockchain development.

\textbf{Takeaway}: Academic research is currently lagging behind industry engagement, but industry efforts are accelerating the development of tools and resources directly aimed at improving BcDEx. This trend points to a clear opportunity for further academic exploration, especially in developing evaluation frameworks specific to blockchain’s environment.

\subsection{In RQ2, we aimed to identify what are the categories of practical sources related to BcDEx}
Our study identified a range of contributions, including tools, frameworks, and guidelines that address distinct aspects of BcDEx. Efficiency in development, usability, and multi-network support emerge as recurring themes across both academic and industry sources~[GL54, GL55, GL06]. In particular, industry publications focus on practical tools and platforms to streamline blockchain development, while academic sources contribute by validating methods and proposing frameworks to enhance BcDEx~[WL04, WL06].

\textbf{Takeaway}: Practical tools for enhancing BcDEx are available and supported by industry initiatives, yet there remains considerable scope for further academic study to systematically validate these contributions. By combining industry-driven innovation with academic rigor, the community could establish more robust BcDEx practices.

\subsection{In RQ3, we explored what ways have the sources discussed in the literature been shaping the BcDEx in practice}
Our study highlights implications for blockchain developers, including enhanced productivity, time savings, and increased adoption of blockchain solutions. Different sources emphasize productivity gains through simplified workflows~[GL09, GL39] and streamlined processes. However, we noticed that few methods or studies exist for specifically evaluating BcDEx, ranging from informal user feedback in industry publications to validation studies in academic literature. This issue highlights an important research gap, as reliable evaluation tools are important for consistently improving BcDEx.

\textbf{Takeaway}: BcDEx has a dual impact, affecting both technical productivity and developer satisfaction. Recognizing and addressing these dimensions is relevant for improving developer retention and fostering sustainable project communities. Developing standardized methods for evaluating BcDEx  could enable more objective and consistent assessment. Future research should prioritize efforts that are specifically suited to blockchain’s technical and social context.

In general, our study emphasizes the importance of BcDEx, noting that industry engagement is advancing practical resources to support developers. Academia has an opportunity to contribute by validating these contributions and by developing evaluation methods. Our findings suggest that a collaborative approach between academia and industry would benefit blockchain ecosystems, making them more developer-friendly.

\section{Threats to Validity}
\label{sec:threats}
We followed the checklist provided by Ampatzoglou et al.~\cite{ampatzoglou2019identifying} to identify threats and their mitigation actions.

\textbf{Study inclusion/exclusion bias.} We identified the completeness of this review could be limited by the scarcity of relevant studies in the WL. To deal with this challenge, we included sources from the GL and applied snowballing techniques (backward and forward) to expand the number of potential primary studies. In addition, we conducted a pilot search to identify terminological variations related to ``developer experience'' in relevant studies.

\textbf{Researcher bias and Repeatability.} We faced challenges related to qualitative coding, which was based only on the textual content of the pages retrieved from the GL. For example, the extraction of features of blockchain development products was performed only from their main pages, potentially omitting details from secondary or internal pages. To mitigate this threat, we followed guidelines from~\cite{garousi2019guidelines} and implemented a quality assessment framework derived from Host and Baysal~\cite{host2007checklists, baysal2023blockchain}. Additionally, a second reviewer performed a thorough review of the extracted data to ensure comprehensive coverage and comparison of key information.

\textbf{Robustness of classification.} We recognize two main threats: the absence of a standardized definition of developer experience'' in the literature and the fact that most sources of GL do not apply rigorous peer review. To reduce the impact of these issues, we adopted Fagerholm and M{\"u}nch's framework~\cite{fagerholm2012developer}, which considers technical aspects, emotional responses, and perception of value in contributions.

\section{Conclusion}
\label{sec:conclusion}

Blockchain development has introduced particular challenges to software engineering, requiring specialized approaches to support developers effectively. Developer Experience (DEx) is a critical factor, as it impacts developers’ productivity, efficiency, and satisfaction. However, studies specifically addressing Blockchain Developer Experience (BcDEx) remain limited, with no systematic mapping exploring how academia and practitioners have approached this topic. This scarcity of studies can hamper efforts to evaluate and improve BcDEx initiatives adequately. In this direction, we conducted a Multivocal Literature Review to analyze the distribution and nature of BcDEx sources, practical implementations, and their impacts shaping this domain in practice. 

Our findings showed that the academic focus on BcDEx is limited compared to the gray literature, which mainly includes blogs and corporate sources focused on improving development efficiency, multi-network support, and usability. Furthermore, we found five perspectives influencing BcDEx in practice: abstracting complexity for better usability, facilitating blockchain adoption, impact on productivity, developer education and support, and BcDEx evaluation. The implications of this research suggest that BcDEx goes beyond improving tools and frameworks and involves rethinking software development in a decentralized environment. Therefore, future work should complement this direction by focusing on developing best practices for BcDEx and establishing effective frameworks for its evaluation. In addition, future research should apply empirical studies to capture individual developers' personal experiences and perceptions directly.

\section*{Acknowledgment}
This work received partial funding from CNPq-Brazil, Universal grant 404406/2023-8, and support from CAPES - Funding Code 001.

\balance
\bibliographystyle{IEEEtran}
\bibliography{main}

\end{document}